\documentclass[preprint,NumberedRefs]{JASAnew}
\usepackage{siunitx} 
\usepackage{blindtext}
\usepackage{float}
\usepackage{upgreek}
\usepackage{graphicx}
\usepackage{booktabs}
\usepackage{wasysym}
\usepackage[utf8]{inputenc}
\usepackage{float}
\usepackage{comment}
\begin{document}
\nolinenumbers
\title{Multi-modal transducer-waveguide construct coupled to a medical needle}

\author{Yohann Le Bourlout}
\affiliation{Medical Ultrasonics Laboratory (MEDUSA), Department of Neuroscience and Biomedical Engineering, Aalto University, Rakentajanaukio 2,  Espoo, 02150, Finland}{}

\author{Gösta Ehnholm}
\affiliation{Medical Ultrasonics Laboratory (MEDUSA), Department of Neuroscience and Biomedical Engineering, Aalto University, Rakentajanaukio 2,  Espoo, 02150, Finland}{}

\author{Heikki J. Nieminen}
\email{heikki.j.nieminen@aalto.fi}
\thanks{Corresponding author}
\affiliation{Medical Ultrasonics Laboratory (MEDUSA), Department of Neuroscience and Biomedical Engineering, Aalto University, Rakentajanaukio 2,  Espoo, 02150, Finland}{}

\preprint{Yohann Le Bourlout, \textit{JASA}}		

\date{\today}

\maketitle

\nolinenumbers
\section*{Abstract}
\nolinenumbers
\noindent Annually, more than 16 billion medical needles are consumed worldwide. However, the functions of the medical needle are still limited to cutting and delivering or drawing material through the needle to a target site. Ultrasound combined with hypodermic needle could potentially add value to many medical applications such as pain reduction, adding precision, deflection reduction in tissues and even improve tissue collection. In this study we introduce a waveguide construct enabling an efficient way to convert a longitudinal wave mode to flexural mode and to couple the converted wave mode to a conventional medial needle, while maintaining high electric-to-acoustic power efficiency. The structural optimization of the waveguide was realized \emph{in silico} using the finite element method followed by prototyping the construct and experimental characterization of the prototypes. The experiments at 30 kHz demonstrate flexural tip displacement up to 200 {\textmu}m, at low electrical power consumption (under 5 W), with up to 69\% of efficiency. The high electric-to-acoustic efficiency and small size of the transducer would facilitate design of medical needle and biopsy devices, potentially enabling portability, batterization and high patient safety with low electric powers.

\newpage

\section{Introduction}
\nolinenumbers
\noindent The hypodermic needle is one of the most commonly used instruments in healthcare, 16 billion needles are consumed annually \cite{Hauri2004TheSettings, Hutin2003UseEstimates}. They are frequently used as standard procedure for tissue sampling \cite{Amedee2001Fine-NeedleBiopsy, Renshaw2007ComparisonBiopsy} or injection of fluids \cite{Cardone2002JointInjection, Wynaden2006BestSetting, Aiello2004EvolvingInjections, Guarda-Nardini2008ArthrocentesisTechnique, Lenz2017AAdministration}. However, only limited development has taken place in their design for several decades\cite{Kirkup1998SurgicalInstruments., Ball2004LocalNeedles}. This has caused complications in some applications, \emph{e.g.} for targeting a specific area for anesthesia or targeting tumor in biopsy \cite{Pritzker2019NeedleCare} (up to 34\% failure rate in breast biopsy using Fine Needle Aspiration (FNA) \cite{Pisano1998RateStudy} and up to 21\% in Thyroid \cite{Gharib1993Fine-NeedleBiopsies}). Moreover, approximatively 10\% of the population fears needles\cite{Nir2003FEARASSOCIATIONS}, which can lead to avoiding accessing healthcare\cite{Hamilton1995NeedleDiagnosis}.

During the operation of the needle, the operator can translate the needle back-and-forth in a piston movement \cite{Kreula1990EffectBiopsy.}, it can be bent to guide the needle and the needle can be moved by rotating it around its center axis by the operator's hand \cite{Han2013StudyBiopsy}. A more modern way to introduce needle movement is to provide mechanical actuation of the needle to produce acoustic waves, essentially providing micron-scale motions within the needle at a rate much greater than possible to produce manually by a human operator. Actuation of the needle tip by ultrasound has recently been demonstrated to be useful for many medical purposes: to (i) decrease the penetration force \cite{Liao2013ReducedStudy, Liao2014PerformanceMimics, Gui2014AutomaticDevice, Yang2004MicroneedleActuation} for an easier and more accurate procedure, (ii) causing less pain by reducing the penetration resistance \cite{Lehtinen1979PenetrationNeedles, Egekvist1999RegionalUltrasonography}, and (iii) provide improved guidance of the biopsy needle under ultrasound imaging \cite{Kuang2016ModellingBiopsy, Sadiq2014EnhancedNeedle, Jiang2020LocalizationImaging}. 

We have recently developed an ultrasound-enhanced fine needle aspiration biopsy (USe-FNAB) method, which allows introducing flexural ultrasound waves to the tip of a medical needle to improve the tissue yield in biopsy \cite{Perra2021UltrasonicYield}. In USeFNAB, a piezoelectric actuator is employed to produce longitudinal waves, which are converted into flexural waves essentially producing mechanical movement of the needle tip. The system previously studied was relatively large in size including a standalone function generator and a power amplifier. However, clinically a small and a fully portable device could be in some circumstances more preferable, requiring high electrical-to-power-conversion efficiency.

In this study, we aimed to develop a new acoustic waveguide construct with the intent to optimize the conversion efficiency from electrical to acoustic power followed by validation of the technical performance.

\section{Material and methods}

\noindent Using the Finite Element (FE) method, we chose to first model the Langevin transducer that was selected as the sound source. This was followed by optimization of the geometry of the full system comprising a piezoelectric transducer, a waveguide and a biopsy needle. Based on the simulations, we then built custom-made prototypes, followed by characterizing their performance and comparing the performance with simulations.

We studied two waveguide constructs: (i) a system with a transducer center axis placed perpendicular to the needle (Figure \ref{PP simu}) and (ii) a system with a transducer aligned at an offset and parallel to the needle (Figure \ref{cuteLsimu}), the needle being a hypodermic needle (21G, length 120 mm, model: 466564/3, 100 Sterican, B Braun, Melsungen, Germany).  The purpose of the waveguide was to convert the longitudinal wave mode in the transducer to a flexural wave mode in the needle, as well as to match the mechanical impedance of the transducer to the mechanical impedance of the needle. The mass distribution, shape and positioning of the structure are critical to achieve optimum performance of the system. Therefore, to understand the wave propagation and optimize the system, an acoustic simulation was conducted \emph{in silico} employing a FE method (COMSOL Multiphysics 5.5., Inc., Burlington, Massachusetts, United States). In the following, the steps realized during this optimization are elaborated.

\subsection{Numerical simulations}

\noindent For optimization of the arrangement, FE method was used (COMSOL Multiphysics 5.5). Electrostatics module was applied to the piezo elements using the stress-charge constitutive relation, associating the electrical field applied to the piezo elements with structural phenomena. Elastodynamics module ("Solid Mechanics" module) was used to solve the equation of motion. 
To avoid infinite vibration amplitude when the system is in resonance, a damping ration of the structural component was set to 0.01.
The different parts of the model were meshed using the swept technique for the piezo stack and the free tetrahedral technique for the rest of the geometry. The mesh size was implemented to include at least 20 elements per wavelength.  
The simulations were conducted in two steps. First, a stationary study was performed to calculate the deformation of the structure, when pretension was applied to the piezo stack as elaborated below. The second step was a frequency domain perturbation study, allowing to simulate the small oscillations of the system.

The parameters of the different materials used in the models are given in the table \ref{table param}.

\subsection{Simulation of the Langevin transducer}

\noindent The modelled Langevin sandwich transducer was composed of twenty-two ring-type piezo elements (PIC151; dimensions = {\diameter}\textsubscript{out} $\times$ {\diameter}\textsubscript{in} $\times$ thickness = 10 $\times$ 5 $\times$ 0.5 mm) stacked and fastened by a bolt. Each element was arranged so that the direction of the electric fields through the piezo were opposite to each other in adjacent elements. Electrically, the elements were all connected in parallel: In this way the required driving amplitude voltage was decreased by a factor of 22 compared to using a single element having a thickness of the full stack. At each end of the stack, an extra inactivated piezo element (PIC151, dimensions = {\diameter}\textsubscript{out} $\times$ {\diameter}\textsubscript{in} $\times$ thickness = 10 $\times$ 5 $\times$ 1 mm) was added to electrically isolate the transducer from the rest of the system for safety. To simulate the glue used by the manufacturer to hold the stack together, two layers of epoxy were added (epoxy, dimensions = {\diameter} $\times$ H = 10 $\times$ 0.2 mm). Two masses of brass were added on each side of the piezo stack connected with a bolt, detailed below, to shift the resonance frequency of the piezo stack to the targeted resonance of the entire construct (approx. 30kHz). The front mass of the transducer was cylindrical (brass, dimensions = H $\times$ {\diameter} = 14 $\times$ 10 mm). The back mass was shaped to facilitate the fixation of the stack inside the enclosure. An M4 screw (brass, H $\times$ {\diameter} = 35 $\times$ 4 mm) was connected to a cylindrical front mass (brass, H $\times$ {\diameter}= 3 $\times$ 10 mm) and to a cube-shaped back mass (brass, H $\times$ W $\times$ L = 10 $\times$ 10 $\times$ 10 mm) connected to a rectangular cuboid (brass, H $\times$ W $\times$ L = 5 $\times$ 10 $\times$ 12 mm). The M4 screw, cylindrical, cube and the rectangular cuboid were all machined from one piece (\emph{c.f.} Figure \ref{langevin transducer}.A).\\
\indent To fasten all the elements mentioned above to create a stack, the M4 bolt linked the back mass and the top mass in order to apply a pretension force (3.1 kN) over the stack. The bolt connects both ends through the top mass via a thread allowing to expose 3 mm of the bolt, which was used to connect the Langevin transducer to the waveguide having a threaded hole. Finally, after a pre-study calculating the bolt pretension effect over the system, a frequency domain with perturbation simulation from 20 to 40 kHz was conducted to find the resonance frequency of the system.

\subsection{Simulations of the waveguides}

\noindent Connecting the needle directly to the transducer without impedance matching would lead to compromised transmission of acoustic power from the transducer to the needle. The transducer stack is vibrating with a longitudinal mode with very small amplitude, and has mechanically a high impedance. On the other hand, the needle should vibrate with a large amplitude approx. perpendicularly to its center axis and is quite flexible, \emph{i.e.} it has a low mechanical impedance. Therefore, mechanical impedance matching is needed.

To avoid reflection of power back to the transducer within the waveguide, a mechanical impedance matching is required. This structure should also convert the longitudinal mode from the transducer to a flexural mode of the needle tip, while matching gradually the transducer mechanical impedance to the needle.
We investigated two waveguide geometries: a linear waveguide and a right-angled waveguide.

The linear waveguide was made of EOS Stainless steel 316L and  included a rectangular resonator in the proximal end (dimensions = L $\times$ W $\times$ H = variable length (8:0.1:100) $\times$ 12 $\times$ 3 mm) (Figure \ref{PP simu}). From here the waveguide extrudes as a tapered section with the width (W) and the height (H) decreasing in width exponentially along the length towards the distal end (L = variable length (5:0.1:50)mm, dimensions starting point = W $\times$ H = 12 $\times$ 3 mm, dimensions ending point = W $\times$ H = 0.9 $\times$ 0.5 mm). The tapering increases the amplitude of the acoustic signal by mechanical impedance matching: The constant flow of power will propagate as a wave with increasing amplitude into more flexible regions at smaller cross sections of the waveguide minimizing reflections. The waveguide also included a groove along the length of the waveguide to provide a coupling area for the needle cannula (dimensions = {\diameter}\textsubscript{in} $\times$ {\diameter}\textsubscript{out}: = 0.514 $\times$ 0.8 mm). 

The second right-angled waveguide design included a linear portion with a 90° angle implemented in the beginning of the tapered portion (Figure \ref{cuteLsimu}). The rationale for this geometry is to improve the ergonomics by aligning the center axes of the transducer and the needle to be parallel. \\
\indent The length dimensions of the waveguides were simulated with the resonance frequency obtained from the transducer simulation.

\subsection{Needle position}
\noindent After optimization of the waveguide, the proximal part of the needle was included into the simulation to understand its effect on the transducer and find an optimal position. The distance from the tip of the waveguide to the needle tip was varied from 8 mm to 60 mm at 100 {\textmu}m increments. The simulated needle follows the specification of a standard 21-gauge (21G) hypodermic needle (dimensions = {\diameter}\textsubscript{in} $\times$ {\diameter}\textsubscript{out} = 0.514 $\times$ 0.8 mm) made of stainless steel 316L. The needle tip, based on standard hypodermic needle tip, is composed of two bevels which have opening angles of 17° and 11°, respectively, observed as a side projection to \emph{y}\emph{z} plane (Figure \ref{PP simu}.C).

\subsection{Impedance characteristics of the experimental system and simulation}
\noindent After building the experimental systems, the agreement between the experimental system and the simulation was verified in terms of impedance characteristics as follows:
The Langevin transducer was coupled to the waveguide, which was a 3D printed EOS stainless steel 316L object (3D Formtech Oy, Jyvaskyla, Finland). The needle 21G x 120 mm (model: 466564/3, 100 STERICAN, B Braun, Melsungen, Germany), placed in the groove, was soldered onto the last 20 mm of the waveguide’s distal end. 
The electrical impedance of the arrangement, was obtained from COMSOL using frequency domain with perturbation simulation (20-40 kHz with steps of 50 Hz). The impedance measurement was conducted using a Digilent Analog Discovery 2 combined with the WaveForms software (Digilent, Inc., Henley Court Pullman, 262 WA, United States). 

\subsection{Quantifying the needle tip displacement}
\noindent To observe the needle tip during sonication we used a high-speed camera (model: Phantom V1612, Vision Research, Wayne, NJ, United States) in conjunction with a macro lens (model: Canon MP-E 65 mm f / 2.8 1-5x Macro Photo, Canon Inc., Ota, Tokyo, Japan). The camera was placed in front of an acrylic chamber (external dimensions = L $\times$ W $\times$ H = 21 $\times$ 14 $\times$ 15 mm, wall thickness = 5 mm) filled with air or deionized water degassed to 6 mg/L at ambient temperature (22.3°C). The needle was immersed to a depth of 30 mm from the water surface. The system was driven by a custom-made amplifier coupled with a custom-made step-up transformer (11:20 turn ratio) at 30.4 kHz, for 10 cycles with an incoming instantaneous power of 5 W. High speed camera videos were taken in order to acquire the needle tip displacement during sonication (sample rate = 304 000 frames per second, exposure time = 0.8{\textmu}s, resolution 128 x 128 pixels, angular magnification = 5x). The videos were then analyzed in MATLAB r2020b using an algorithm capable of tracking the needle position using cross-correlation frequency technique based on Perra et al. \cite{Perra2021UltrasonicYield} work.

\subsection{Calorimetry}
\noindent To define the total acoustic power emitted from the needle, a calorimetric assessment was done. The calorimeter (material= thermal wall insulator, wall thickness= 3 cm) was filled with 25 mL of deionized water (at room temperature: 22°C) mixed with a magnetic stirrer (C-MAG HS series C-MAG HS 4 model, IKA, Staufen, Germany). The needle tip was then immersed to a depth of 30 mm from the water surface. The system was driven at 30.4 kHz using a burst signal with 55\% activation time (used for a better control at the needle tip) during 10 seconds. The forward and reflected powers were measured and time-averaged over a burst period: The time-averaged consumed power is equal to their difference. The water temperature change was recorded using a PT 3000 thermometer probe for 10 seconds before and 10 seconds after the sonication. After averaging the temperature data, the difference of the two temperatures gave the heat deposed into a known volume of distilled water for a certain incident energy. The analysis of the data was implemented using a custom-made MATLAB r2020b software.

\section{Results}
\subsection{Langevin transducer}
\begin{figure}[H]
\begin{center}
\includegraphics[width=\textwidth]{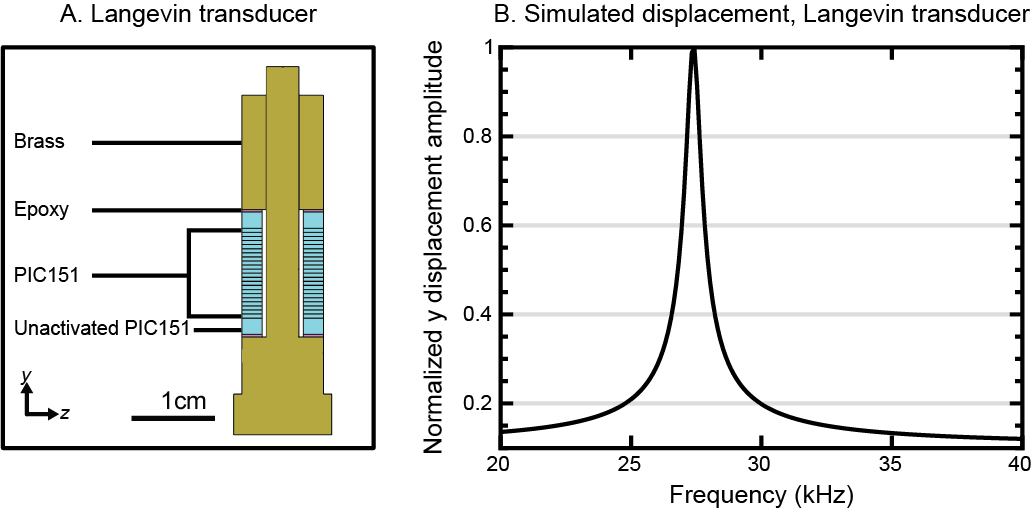}
\caption{A. Cross-section of the Langevin transducer construct with twenty-two rings arranged so that their electric fields direction through the piezo were opposite with adjacents piezo elements. At each end of the piezo stack, an extra non-activated piezo element was placed to isolate the transducer from the rest of the system for safety. At the extremities of the transducer, two masses made of brass were included. An M4 bolt connecting the two masses were applied to prestress the piezo stack. B. Absolute displacement measured from a frequency domain simulation using COMSOL multiphysics 5.5 in \emph{y} direction with 1 W driving power.}
\label{langevin transducer}
\end{center}
\end{figure}

\noindent The first analysis aimed to define the resonance frequency of the transducer. The computational results indicate a peak vibration amplitude around 27.4 kHz as shown in Figure \ref{langevin transducer}.

\subsection{Optimization of the resonator, linear waveguide and needle length}

\begin{figure}[H]
\includegraphics[width=\columnwidth]{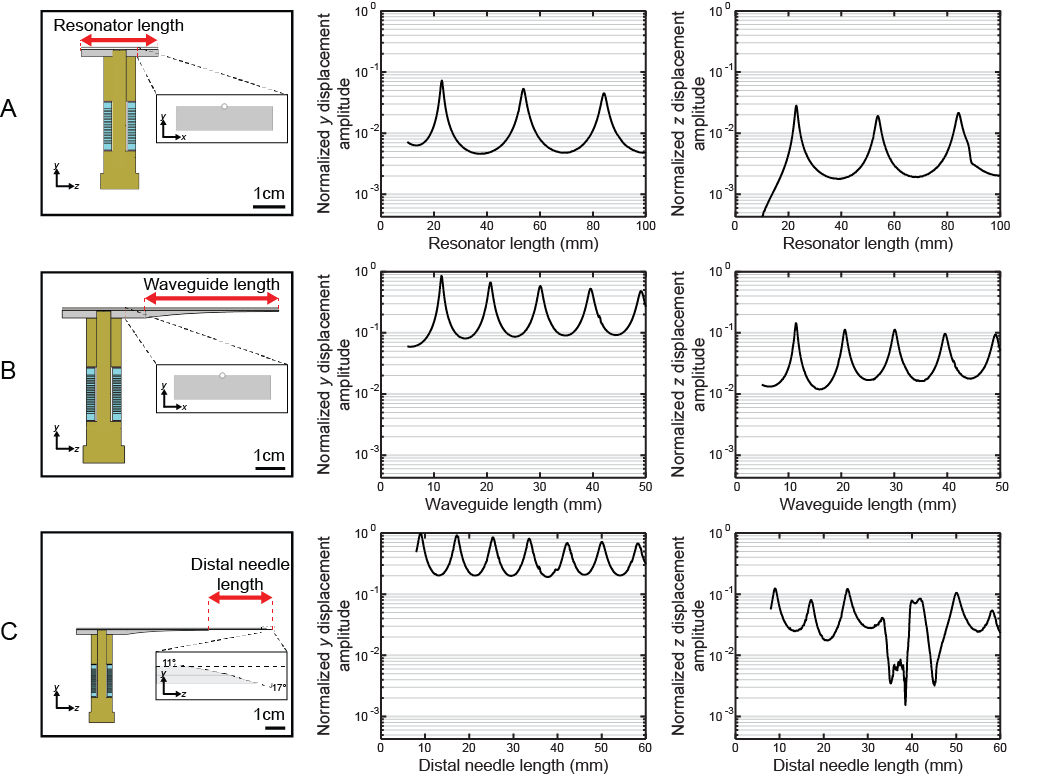}
\caption{A. Displacement in \emph{y} and \emph{z} direction dependent of the resonator length variation. B. Displacement in \emph{y} and \emph{z} direction dependent of the amplifier length variation. C. Displacement in \emph{y} and \emph{z} direction dependent of the needle length variation}
\label{PP simu}
\end{figure}

\noindent The purpose of this simulation was to find a geometry of the waveguide, which would provide a great displacement in \emph{y} direction while avoiding ‘spurious’ displacement in other directions due to different modes (As shown in the row A of the normalized \emph{z} displacement graph around 82 mm). In Figure \ref{PP simu}.A, there is a clear recurrence, every 23 mm of the greater values which corresponds to approx. half a wavelength. The selected dimension was 23 mm. The same applies in Figure \ref{PP simu}. B and C, every half a wavelength, an optimum in \emph{y} displacement appearing. 39.5 mm was selected as the length of the waveguide. The displacement is greatly amplified by the tapered shape compared to the simple resonator. The needle displacement is also, greater than the amplifier construct. This is due to the tapered proximal end of the needle which behaves as an amplifier. The distance distal part of the needle to distal part of the waveguide was 58 mm. 

\subsection{Optimization of the resonator, right-angled waveguide and needle length}

\begin{figure}[H]
\includegraphics[width=\columnwidth]{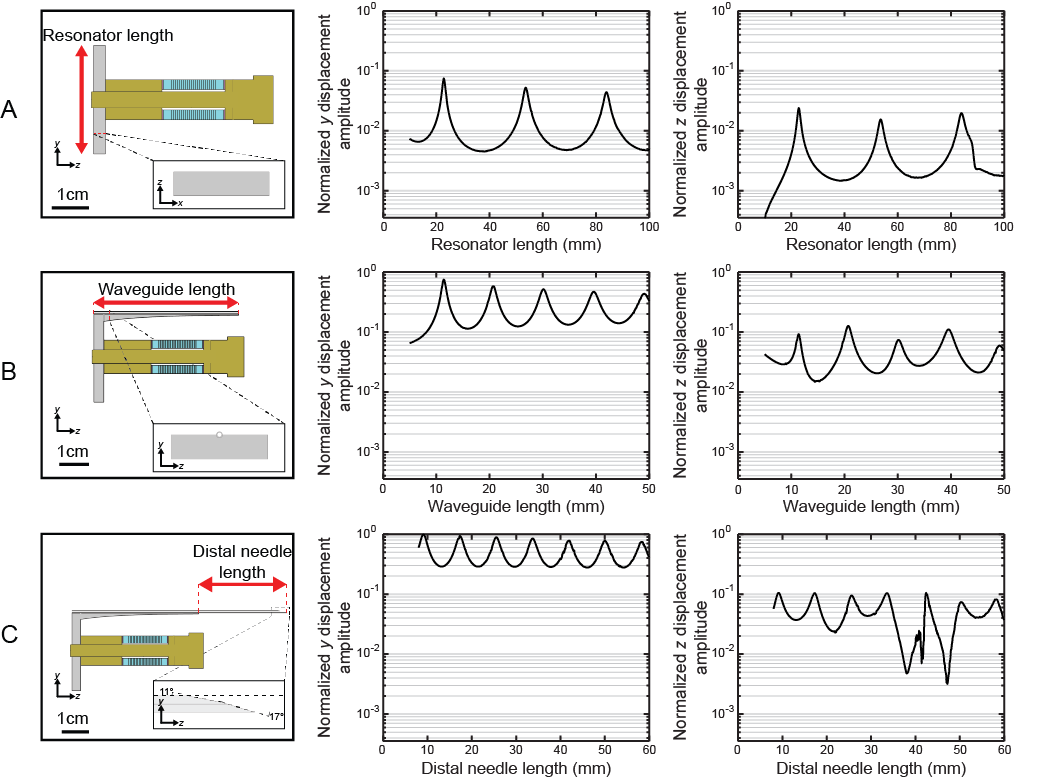}
\caption{A. Displacement in \emph{y} and \emph{z} direction dependent of the resonator length variation. B. Displacement in \emph{y} and \emph{z} direction for the right-angled design dependent of the amplifier length variation. C. Displacement in \emph{y} and \emph{z} direction for the right-angled design dependent of the needle length variation.}
\label{cuteLsimu}
\end{figure}

\noindent The results show in the Figure \ref{cuteLsimu}, using the right-angled waveguide suggest a similar functioning as the linear waveguide. For the optimization of the amplifier, a resonator of half a wavelength was selected (23 mm). After the simulation of the amplifier, a length of 39.5 mm was selected for the compromise between length of the waveguide (long enough to reduce the reflection of the tapering) and the useful needle length (greater the waveguide length is, smaller is the proximal size of the needle). The proximal length of the needle selected for this model was 58 mm.

\subsection{Validation model - system}

\begin{figure}[H]
\includegraphics[width=\textwidth]{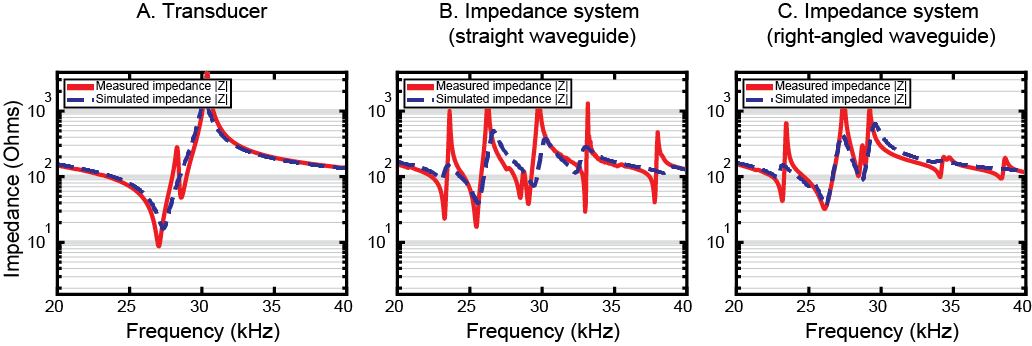}
\caption{A. Comparison between the simulated impedance of the transducer construct and the measured impedance of the custom-designed transducer. B. Comparison of the whole system impedance including the transducer, linear waveguide and needle simulation and experimental device with linear waveguide. C. Comparison of the whole system with right-angled waveguide and needle simulation and experimental device with right-angled waveguide.}
\label{fig4}
\end{figure}

\noindent To compare the difference between the simulations and our prototypes, an impedance measurement with bare stack transducer and the whole systems were compared with the respective simulated impedances in frequency range 20 - 40 kHz (Figure \ref{fig4}). The simulation curve and the measured peaks exhibit similar behavior.
The main impedance peak from the transducer is at 30.3 kHz, while the linear waveguide system seems to be at 29.7 kHz and the right-angled one appears to be at 29.1 kHz, which remains in the desired frequency range.

\subsection{Experiment arrangement}

\begin{figure}[H]
\includegraphics[width=\textwidth]{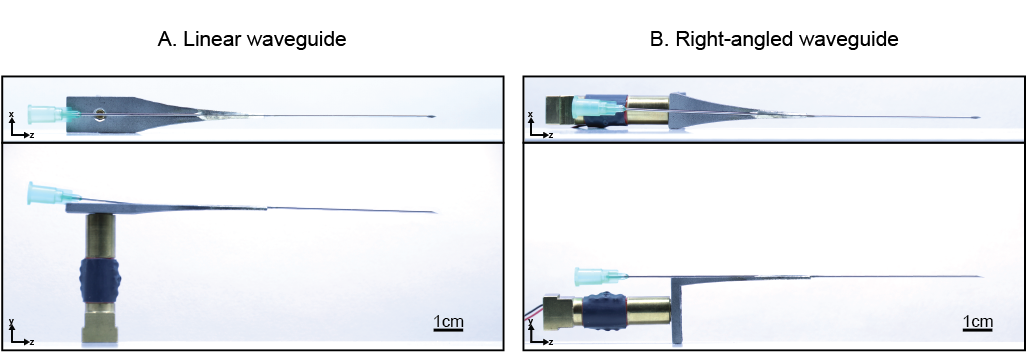}
\caption{Pictures of the two concepts. A. Shows the top and side view of the device with the linear waveguide, while B. shows the device with right-angled waveguide. Here the pictures exhibit the more compact size of the right-angled waveguide.}
\label{devices}
\end{figure}

\begin{figure}[H]
\includegraphics[width=\textwidth]{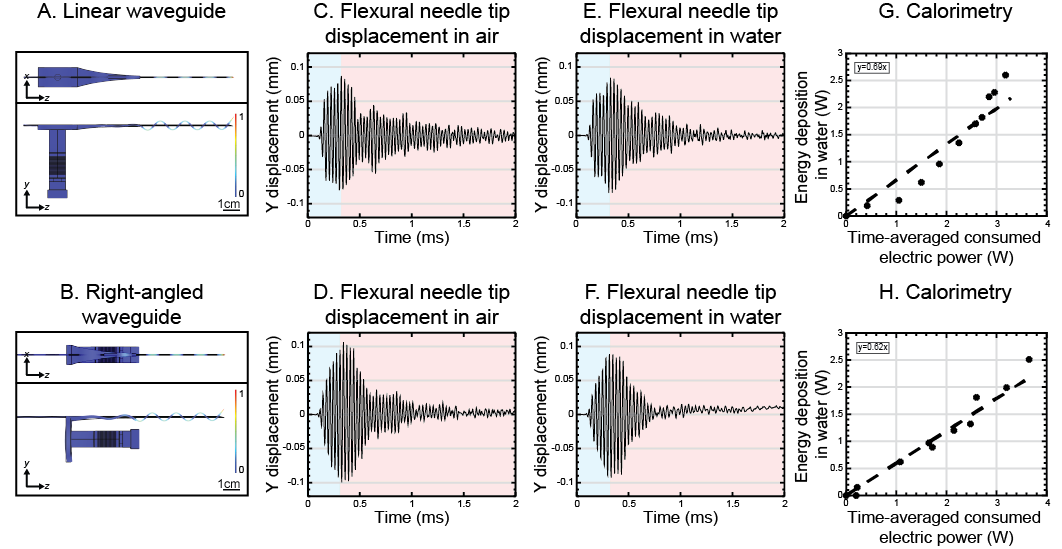}
\caption{Frequency domain simulation (COMSOL multiphysics 5.5) of the linear (A) and right-angled (B) waveguides (an exaggerated deformation has been used for better visualization), their experimental displacement of the needle in air (C,D) or water (E,F) (30.4kHz, 10 cycles, instantaneous electrical power 5W, in blue background) recorded with a high speed camera (304000 fps) and their needle tip heating deposition in water measurement (C,F) respectively.}
\label{experiments}
\end{figure}

\noindent Figure \ref{devices} shows the different prototypes. Figure \ref{experiments} compares the waveguide structures and shows that both of the designs are able to create flexural waves at the needle tip (Figure \ref{experiments}.A-F). Both systems are behaving similarly (Figure \ref{experiments}.C-F), a needle tip displacement up to 200 {\textmu}m and 160{\textmu}m, in air and water, respectively. Cavitations were observable with both waveguides from the high speed camera footage in water. Nevertheless, it seems that a slightly greater electrical-to-acoustical power -efficiency is obtained with the linear waveguide (Figure \ref{experiments}.G,H): 69\% efficiency was recorded with the linear waveguide compared to an efficiency of 62\% with the right-angled waveguide.

\section{Discussions}
\noindent The results of the \emph{in silico} structural optimization demonstrated the displacements at the needle tip were significantly dependent on the dimensions of the waveguide design and needle-to-waveguide coupling point as expected, greatest displacements obtained when the system was in resonance. In the simulations, an increase in displacement was observed at the end of each stage, \emph{e.g.} transducer, waveguide resonator or waveguide tapering, to the end of next system stage, i.e. waveguide resonator, waveguide tapering or needle, respectively. Importantly, the experimentally demonstrated power-efficiency was good, 69\% with the linear waveguide, 62\% with the right-angled waveguide. Several factors may contribute to the power-efficiency, as outlined in the following. 

The extrusion of the linear and rectangular segment of the resonator extends symmetrically from the transducer in opposite directions. Symmetricity minimizes the generation of bending modes in the transducer at the coupling point to the resonator, which would be more likely with an asymmetric resonator. Bending waves in the transducer would yield in tension and compression in different parts of the piezo stack simultaneously, compromising the performance, which is now eliminated. Furthermore, the exponential tapered shape of the waveguide gives an amplification of the signal by having approximately the same energy in the smaller cross-section, while minimizing the reflection from the boundaries towards the source. The geometry, therefore, provides geometric amplification at the same time with successful impedance matching from higher mechanical impedance of the transducer to the lower mechanical impedance of the needle. The needle bevel provides further an additional state of displacement amplification \cite{Perra2021UltrasonicYield}.

It is notable that the studied waveguides converted the longitudinal waves emitted by the transducer to a flexural displacement of the needle tip. The experiments with needle in air provided large displacement amplitudes up to 200 {\textmu}m and 160 {\textmu}m in water. The generation of cavitation was observed using both the linear and right-angled waveguides. The practical relevance of the potential to generate large displacements is selecting the relevant magnitude needed for a specific application in question. \emph{E.g.} gentle actuation of tissue by the needle or adjacent micro-bubbles could be used to improve tissue yield in a biopsy application, whereas generation of large amplitudes and violent cavitation clouds could be useful in an intervention application such as histotripsy of a tumor.

Several reports \cite{Liao2013ReducedStudy, Liao2014PerformanceMimics} have shown that the actuation of the needle tip by ultrasound might be useful for many medical purposes such as pain reduction or improving spatial accuracy of needle insertion and avoiding the tissue deformation. Some devices have met these expectations in terms of needle tip actuation, but none have been small enough to be adopted by doctors. An initial objective of the project was to create an ultrasonic actuated needle with high efficiency system. This could permit batterization and miniaturization if developed to its full potential. The instantaneous electrical input power into the transducer was below 5 W and the displacements shown here are sufficient in the biopsy application \cite{Perra2021UltrasonicYield}, which could be feasible to provide in a miniaturized fashion.

The system could be further improved with a resonance frequency tracking, which was not implemented in this study. The loading of the needle affects the resonance frequency. Increasing the load of the needle leads to a decrease of the resistive power and an increase of the reactive power, which would decrease the efficiency of the system. The frequency tracking for finding the resonance frequency could thereby enable minimizing reactance and thereby maintaining maximal needle tip action despite the mechanical environment of the needle. The mechanical environment would change \emph{e.g.} with the needle penetration depth in tissue or with different mechanical properties of the tissue. In our experiment, the effect of the water on the frequency change was not investigated, but stiffer media could have a greater effect on the frequency, leading to a non-optimal system.

\section{Conclusion}
\noindent Ultrasound in medicine is a well known technology and is widely used for different kind of therapeutic procedures. However, mechanical ultrasound in healthcare is not widespread. Since the needle is a common tool in healthcare but with clear limitations, this paper showed a novel way of improving needles. By actuating the needle with a flexural wave, we might be able to decrease the penetration force needed during the insertion of the needle or improve the tissue collection during biopsies. Due to the efficiency of the system, it suggests that the design could lead to a batterization and, eventually, enhancing the miniaturization of the system.

\section{Acknowledgment}
\noindent We thank Mr. Emanuele Perra, M.Sc., Mr. Matti Mikkola, M.Sc., Mr. Saif Bunni, M.Sc., Mr. Jouni Rantanen, M.Sc. and Dr. Nick Hayward, MBBS, Ph.D. for constructive discussions related to the topic. Business Finland and Academy of Finland are acknowledged for financial support (Business Finland: grant 5607/31/2018; Academy of Finland: grants 314286, 311586 and 335799). We thank also Alto University, the Department of Neuroscience and Biomedical Engineering and MEDUSA laboratory for the facilities and equipment.

\section{Conflict of interest}
\noindent Heikki Nieminen has stock ownership in Swan Cytologics Inc., Toronto, ON, Canada and is an inventor within the patent application WO2018000102A1. Yohann Le Bourlout, Gösta Ehnholm and Heikki Nieminen are inventors within the patent application PCT/FI2020/050345.

\section{Contributions}
\noindent All authors contributed to the design of the study, writing or reviewing the manuscript and have approved it. Yohann Le Bourlout have produced the data and conducted the data analysis. Gösta Ehnholm have assembled the transducer construct and created the electronics for the system.

\section{Data availability}
\noindent The data is available upon request.

\section{Code availability}
\noindent The codes are available upon request.

\bibliography{referencesMendeley.bib}
\pagebreak
\section{Annexes}

\begin{table}[htb]
\centering
\caption{Material parameters used for the FE simulations}
\label{table param}
\begin{tabular}{|c|c|c|c|c|}
\hline
Material & Density [kg/m$^{3}$] & Young's modulus [GPA] & Poisson's ratio & Isotropy\\ [0.5ex] 
\hline
Brass & 8730 & 82 & 0.31 & isotropic \\ 
\hline
Epoxy & 1250 & 3.5 & 0.3 & isotropic\\
\hline
PIC151 & 7760 & Manufacturer data sheet & Manufacturer data sheet & anisotropic\\
\hline
EOS Stainless steel 316L & 8000 & 193 & 0.265 & isotropic\\
 \hline
Stainless steel 316L & 8070 & 205 & 0.275 & isotropic\\
 \hline
\end{tabular}
\end{table}

\end{document}